\documentclass[showpacs,showkeys]{revtex4}
\usepackage{graphicx}
\usepackage{epstopdf}

\usepackage{amsmath}
\usepackage{amssymb}
\usepackage{color}
\usepackage{yfonts}

\begin{document}

\title{
Topologically nontrivial solution in Einstein-Dirac gravity on the Hopf bundle
}

\author{Vladimir Dzhunushaliev}
\email{v.dzhunushaliev@gmail.com}
\affiliation{
	Dept. Theor. and Nucl. Phys., KazNU, Almaty, 050040, Kazakhstan \\
	IETP, Al-Farabi KazNU, Almaty, 050040, Kazakhstan \\
	Institute of Systems Science,
	Durban University of Technology, P. O. Box 1334, Durban 4000, South Africa
}

\date{\today}

\begin{abstract}
The topologically nontrivial solution in Einstein-Dirac gravity with cosmological constant is obtained. The spacetime has the Hopf bundle as a spatial section. It is shown that the Hopf invariant is related to the spinor current density. Two Dirac spinors are used for obtaining a diagonal energy-momentum tensor. The solutions for the nongravitating Dirac
equation on the background of Lorentzian spacetime with the Hopf bundle as a spatial section are also obtained. Nongravitating solutions of the Dirac equation are defined by two quantum half-integer numbers $m, n$.
\end{abstract}

\pacs{}
\keywords{Einstein-Dirac gravity, Hopf bundle, topologically nontrivial gravitating solution, nongravitating solutions}

\maketitle

\section{Introduction}

In general relativity, there are a lot of solutions with gravitating fundamental fields:  scalar, electromagnetic, and non-Abelian fields.
However, very little is known about solutions in Einstein-Dirac gravity. For example, we do not know any asymptotically flat solution
with a gravitating spinor field. Perhaps the problem here is related  to the fact that the spinor field has a spin.
This has the result
that the energy-momentum tensor for a spinor field has nondiagonal components and not only
$T_{t \varphi}$, as it happens for the Kerr metric.

Gravitating spinors with nonlinear self-interactions are well investigated in cosmology \cite{Saha:2016cbu}-\cite{Saha:2014hea}.
These papers study the role of a spinor field in considering the evolution of the
anisotropic Universe described by the Bianchi type  VI, VI$_0$, V, III, I, or isotropic Friedmann-Robertson-Walker (FRW) models.
In Refs.~\cite{Ribas:2016ulz} and \cite{Ribas:2010zj} models of the Universe with tachyonic and fermionic fields interacting
through a Yukawa-type potential are investigated. In Ref. \cite{Balantekin:2007km} a
class of exact cosmological solutions with a neutral scalar field and a Majorana fermion field is found. A Dirac spinor in $D = 3$
dimensions coupled to topologically massive gravity is investigated in \cite{Adak:2004jk}. In Ref. \cite{Finster:2008hc}
a mechanism where quantum oscillations of the Dirac wave functions can prevent the formation of the big bang or big crunch singularity is analysed.

A simpler problem is the problem of looking for solutions on the background of a curved spacetime. In the textbook \cite{Chandrasekhar:1985kt}
Dirac's equation on the background of Kerr geometry is considered. In Ref. \cite{Goatham:2009ek} topologically trivial solutions for the Dirac equation on a 3D sphere $S^3$ are obtained.

Here we will consider two topologically nontrivial Dirac spinors coupled to Einstein gravity with the cosmological constant. 3D section of the spacetime metric is a Hopf bundle.
The topological nontriviality means that the current of a spinor field is connected with the Hopf invariant. We consider two Dirac spinors.
The energy-momentum tensor for every spinor field has nondiagonal components that is related to the fact that the spinor has the spin.
For our choice of spinors the total energy-momentum tensor will have diagonal components only.

We will also investigate the nongravitating Dirac equation. Some special solutions will be found for the case when a Dirac spinor
can be decoupled on Weyl spinors (this is the case when the mass of a spinor is zero, $m=0$). Numerical solution for non-zero mass $m \neq 0$ will be also presented.

\section{Einstein-Dirac gravity}

Einstein-Dirac equations for two gravitating spinors field $\psi_{1,2}$ are
\begin{eqnarray}
	R_{a b} - \frac{1}{2} \eta_{a b} R - \eta_{ab} \Lambda 
	&=& \varkappa T_{a b} ,
\label{0-10}\\
	\left(
		i \gamma^\mu D_\mu - \tilde \mu_{1,2}
	\right) \psi_{1,2} &=& 0 .
\label{0-20}
\end{eqnarray}
Here Latin letters $a, b = 0,1,2,3$ are related to vielbein indices; Greece letters $\mu = 0,1,2,3$ are spacetime indices;
$D_\mu = \partial_\mu + \frac{1}{4} \omega_{a b \mu} \gamma^a \gamma^b$
is the spinor covariant derivative; $\omega_{a b \mu}$ is the spin connection; the energy-momentum tensors for two spinor fields $\psi_{1,2}$ are
\begin{eqnarray}
	T_{\mu \nu} &=& T_{1; \mu \nu} + T_{2; \mu \nu} ,
\label{0-30}\\
	T_{1,2; \mu \nu} &=& \frac{i}{4} \left(
		\bar \psi_{1,2} \gamma_\mu D_\nu \psi_{1,2} - D_\nu \bar \psi_{1,2} \gamma_\mu \psi_{1,2} +
		\bar \psi_{1,2} \gamma_\nu D_\mu \psi_{1,2} - D_\mu \bar \psi_{1,2} \gamma_\nu \psi_{1,2}
	\right)
\label{0-50}
\end{eqnarray}
The spin connection $\omega_{a b \mu}$, the Ricci coefficients $\Delta_{\alpha \beta \gamma}$,
the anholonomy coefficients $\Sigma^a_{\phantom{a} \mu \nu}$ are defined as (here we follow the textbook \cite{poplawski})
\begin{eqnarray}
	\omega_{a b \mu} &=& - e_a^{\phantom{a} \alpha}
	e_b^{\phantom{b} \beta} \Delta_{\alpha \beta \mu} ,
\label{1-31}\\
	\Delta_{\alpha \beta \gamma} &=&
	e_{a \alpha} \Sigma^a_{\phantom{a} \beta \gamma} -
	e_{a \beta} \Sigma^a_{\phantom{a} \alpha \gamma} -
	e_{a \gamma} \Sigma^a_{\phantom{a} \alpha \beta},
\label{0-60}\\
	\Sigma^a_{\phantom{a} \mu \nu} &=&
	\frac{1}{2} \left(
	\partial_\nu e^a_{\phantom{a} \mu} -
	\partial_\mu e^a_{\phantom{a} \nu}
	\right) .
\label{0-70}
\end{eqnarray}
Dirac matrices $\gamma^a$ in flat space are
\begin{equation}
	\gamma^a  = \left\lbrace
		\begin{pmatrix}
			0 & 1 \\
			1 & 0
		\end{pmatrix},
		\begin{pmatrix}
			0 		 & - \sigma^j \\
			\sigma^j & 0
		\end{pmatrix}
	\right\rbrace  , j = 1,2,3,
\label{0-75}
\end{equation}
where $\sigma^j$ are the Pauli matrices
\begin{equation}
	\sigma^j  = \left\lbrace
	\begin{pmatrix}
		0 & 1 \\
		1 & 0
	\end{pmatrix},
	\begin{pmatrix}
		0 & - i \\
		i & 0
	\end{pmatrix},
	\begin{pmatrix}
		1 & 0 \\
		0  & - 1
	\end{pmatrix}
	\right\rbrace .
\label{0-80}
\end{equation}

\section{The solution}

We seek the solution of the Einstein-Dirac equations \eqref{0-10} and \eqref{0-20} in the following form:
\begin{eqnarray}
	ds^2 &=& dt^2 - \frac{r^2}{4} \left[
	\left( d\chi^2 - \cos \theta d \varphi \right)^2
	+ d \theta^2 + \sin^2 \theta d \varphi^2
	\right] ,
\label{1-10}\\
	\psi_1 &=& e^{i \tilde \Omega_1 t} e^{i n_1 \chi} e^{i m_1 \varphi} \begin{pmatrix}
		\Theta_1(\theta) & \\
		\Theta_2(\theta) & \\
		\Theta_3(\theta) & \\
		\Theta_4(\theta) &
	\end{pmatrix} ,
	\psi_2 = e^{i \tilde \Omega_2 t} e^{i n_2 \chi} e^{i m_2 \varphi} \begin{pmatrix}
		\Sigma_1(\theta) & \\
		\Sigma_2(\theta) & \\
		\Sigma_3(\theta) & \\
		\Sigma_4(\theta) &
	\end{pmatrix},
\label{1-60}
\end{eqnarray}
where the space metric
\begin{equation}
	dl^2 = \frac{r^2}{4} \left[
		\left( d\chi^2 - \cos \theta d \varphi \right)^2
		+ d \theta^2 + \sin^2 \theta d \varphi^2
	\right]
\label{1-15}
\end{equation}
is the metric on the Hopf bundle, and $r$ is the radius of 3D sphere. The Hopf bundle can be presented as a $S^3$ sphere with topological mapping
$S^3 \rightarrow S^2$,
where the fibre $S^1$ is spanned on the coordinate $\psi$, and the metric on the base of bundle $S^2$ is
$dl^2 = ( r/2 )^2 \left(
d \theta ^2 + \sin^2 \theta d \varphi^2
\right) $.

In order to calculate all these quantities, we have to define tetrads $e^a_{\phantom{a} \mu}$ for the metric \eqref{1-10}:
\begin{equation}
	e^a_{\phantom{a} \mu} =
	\begin{pmatrix}
		1 & 		0 		& 0 			& 0 						\\
		0 & 	\frac{r}{2}	& 0 			& -\frac{r}{2} \cos \theta 	\\
		0 & 0				& \frac{r}{2} 	& 0 						\\
		0 & 0				& 0 			& \frac{r}{2} \sin \theta	
	\end{pmatrix} .
\label{1-35}
\end{equation}

\subsection{Dirac equations on the Hopf bundle}

The spinors \eqref{1-60} may transform under the rotation at an angle $2 \pi$  as follows:
\begin{eqnarray}
	\psi_{1,2}(\chi + 2 \pi) &=& \pm \psi_{1,2}(\chi),
\label{1-120}\\
	\psi_{1,2}(\varphi + 2 \pi) &=& \pm \psi_{1,2}(\varphi).
\label{1-140}
\end{eqnarray}
Taking into account the exponents $e^{i n_{1,2} \chi}, e^{i m_{1,2} \varphi}$ from \eqref{1-60}, we see that the numbers $m_{1,2}, n_{1,2}$ should satisfy the following condition:
\begin{equation}
	m_{1,2} + n_{1,2} = \pm \frac{N}{2},
\label{1-160}
\end{equation}
where $N$ is an integer.

After substitution the $\mathfrak{Ansatz}$ \eqref{1-60} into the Dirac equation \eqref{0-20}, we have
\begin{eqnarray}
	\Theta_1^\prime + \Theta_1 \left(
		\frac{\cot \theta}{2} + n
	\right) +
	\Theta_2  \left(
		\frac{1}{4} - \frac{\Omega}{2} - n \cot \theta - \frac{m}{\sin \theta}
	\right) - \frac{\mu}{2} \Theta_4
	&=& 0 ,
\label{1-80}\\
	\Theta_2^\prime + \Theta_2 \left(
		\frac{\cot \theta}{2} - n
	\right) +
	\Theta_1  \left(
		- \frac{1}{4} + \frac{\Omega}{2} - n \cot \theta - \frac{m}{\sin \theta}
	\right) + \frac{\mu}{2} \Theta_3
	&=& 0 ,
\label{1-90}\\
	\Theta_3^\prime + \Theta_3 \left(
		\frac{\cot \theta}{2} + n
	\right) + \Theta_4
	\left(
		\frac{1}{4} + \frac{\Omega}{2} - n \cot \theta - \frac{m}{\sin \theta}
	\right) + \frac{\mu}{2} \Theta_2
	&=& 0 ,
\label{1-100}\\
	\Theta_4^\prime + \Theta_4 \left(
		\frac{\cot \theta}{2} - n
	\right) + \Theta_3
	\left(
		- \frac{1}{4} - \frac{\Omega}{2} - n \cot \theta - \frac{m}{\sin \theta}
	\right) - \frac{\mu}{2}\Theta_1
	&=& 0 ,
\label{1-110}
\end{eqnarray}
where, for brevity, we have omitted the index: $\Omega = \Omega_1$, $m,n = m_1 , n_1$; $\Omega = r \tilde \Omega$ and $\mu = r \tilde \mu$.
For $\Sigma_{1,2,3,4}$ we also have the same equations. Some special solutions for this set of equations  are presented in Appendixes \ref{app1} and \ref{app2}.

\subsection{Energy-momentum tensor for massless spinor fields, $m=0$}

In order to solve the Einstein-Dirac equations \eqref{0-10} and \eqref{0-20}, we will use the following
\textgoth{Ans\"atze} for the spinors $\psi_{1,2}$:
\begin{eqnarray}
	\psi_{1,2} &=& e^{i \tilde \Omega_{1,2} t} e^{i \chi/2}
	\tilde \Theta_{1,2} \begin{pmatrix}
		1 & \\
		1 & \\
		0 & \\
		0 &
	\end{pmatrix} ,
	T_{1,2; a b} = \frac{\tilde \Theta^2_{1,2}}{r}
	\begin{pmatrix}
		3 	& -2	&	0	&	0	\\
		-2	& 1		&	0	&	0	\\
		0	& 0		&	1	&	0	\\
		0	& 0		&	0	&	1	
	\end{pmatrix} ,
	\tilde \Omega_{1,2} = \frac{3}{2 r} ;
\label{3b-10}\\
	\psi_{1,2} &=& e^{i \tilde \Omega_{1,2} t} e^{- i \chi/2}
	\tilde \Theta_{1,2} \begin{pmatrix}
		\phantom{-}1 & \\
		-1 & \\
		\phantom{-}0 & \\
		\phantom{-}0 &
	\end{pmatrix} ,
	T_{1,2; a b} = \frac{\tilde \Theta^2_{1,2}}{r}
	\begin{pmatrix}
		 3 	& 2		&	0	&	0	\\
		2	& 1		&	0	&	0	\\
		0	& 0		&	1	&	0	\\
		0	& 0		&	0	&	1	
	\end{pmatrix} ,
	\tilde \Omega_{1,2} = \frac{3}{2 r} ;
\label{3b-20}\\
	\psi_{1,2} &=& e^{i \tilde \Omega_{1,2} t} e^{i \chi/2}
	\tilde \Theta_{1,2} \begin{pmatrix}
		0 & \\
		0 & \\
		1 & \\
		1 &
	\end{pmatrix} ,
	T_{1,2; a b} = \frac{\tilde \Theta^2_{1,2}}{r}
	\begin{pmatrix}
		-3 	& -2	&	0	&	0	\\
		-2	& -1	&	0	&	0	\\
		0	& 0		&	-1	&	0	\\
		0	& 0		&	0	&	-1	
	\end{pmatrix} ,
	\tilde \Omega_{1,2} = - \frac{3}{2 r} ;
\label{3b-30}\\
	\psi_{1,2} &=& e^{i \tilde \Omega_{1,2} t} e^{- i \chi/2}
	\tilde \Theta_{1,2}	\begin{pmatrix}
		\phantom{-}0 & \\
		\phantom{-}0 & \\
		\phantom{-}1 & \\
		-1 &
	\end{pmatrix} ,
	T_{1,2; a b} = \frac{\tilde \Theta^2_{1,2}}{r}
	\begin{pmatrix}
		-3 	& 2	&	0	&	0	\\
		2	& -1&	0	&	0	\\
		0	& 0	&	-1	&	0	\\
		0	& 0	&	0	&	-1	
	\end{pmatrix} ,
	\tilde \Omega_{1,2} = - \frac{3}{2 r}.
\label{3b-40}
\end{eqnarray}
Here $\tilde \Theta_{1,2}$ are constants.

\subsection{The solution of the Einstein-Dirac equations}

Einstein tensor $G_{ab} = R_{ab} - (1/2) \eta_{ab} R$ for the tetrad \eqref{1-35} is
\begin{equation}
	G_{a b} = \frac{1}{r^2} \text{diag} \left(
		3, -1, -1, -1
	\right) .
\label{3c-05}
\end{equation}
In order to have the same structure on the right-hand side of the Einstein equations [as in \eqref{3c-05}], we have to use the sum of energy-momentum tensors from \eqref{3b-10} and \eqref{3b-20}. In this case the Einstein equations will be
\begin{eqnarray}
	\frac{3}{r^2} - \Lambda &=& 48 \pi l^2_{Pl} 
	\frac{\tilde \Theta^2}{r} ,
\label{3c-10}\\
	- \frac{1}{r^2} + \Lambda &=& 16 \pi l^2_{Pl} \frac{\tilde \Theta^2}{r} 
\label{3c-15}
\end{eqnarray}
here $\tilde \Theta = \tilde \Theta_{1,2}$ and $l_{Pl}$ is the Planck length. The solution of these equations in dimensionless form is
\begin{eqnarray}
	\tilde \Theta l^{3/2}_{Pl} &=& \left(
		\frac{1}{32 \pi} \frac{l_{Pl}}{r}
	\right)^{1/2} ,
\label{3c-20}\\
	\Lambda &=& \frac{3}{2 r^2} .
\label{3c-30}
\end{eqnarray}
It is useful to give some numerical estimation for this solution. For example, if we take $r \approx \Lambda^{-1/2}_{obs} \approx 10^{26}$m, then 
\begin{equation}
	\tilde \Theta l^{3/2}_{Pl} \approx \left( 
		\frac{1}{32 \pi} l_{Pl} \Lambda^{1/2}_{obs} 
	\right) \approx 10^{-31}
\label{3s-25}
\end{equation}
here $\Lambda_{obs} $ is observed value of the cosmological constant. 

Finally, the solution of the Einstein-Dirac equations \eqref{0-10} and \eqref{0-20} is
\begin{eqnarray}
	ds^2 &=& dt^2 - \frac{r^2}{4} \left[
	\left( d\chi^2 - \cos \theta d \varphi \right)^2
	+ d \theta^2 + \sin^2 \theta d \varphi^2
	\right] ,
\label{3c-60}\\
	\psi_1 &=& e^{i \tilde \Omega t} e^{i \chi/2} \tilde \Theta
	\begin{pmatrix}
		1 & \\
		1 & \\
		0 & \\
		0 &
	\end{pmatrix} ,
\label{3c-70}\\
	\psi_2 &=& e^{i \tilde \Omega t} e^{- i \chi/2} \tilde \Theta
	\begin{pmatrix}
		\phantom{-}1 & \\
		-1 & \\
		\phantom{-}0 & \\
		\phantom{-}0 &
	\end{pmatrix} , \tilde \Omega = \frac{3}{2 r}
\label{3c-80}
\end{eqnarray}
with $\tilde \Theta$, $\epsilon$ from \eqref{3c-20} and \eqref{3c-30}, and $r$ is the radius of a 3D sphere that is the total space of the Hopf bundle.

\subsection{Current and the Hopf invariant}

The covariant current for the spinor $\psi_{1}$ is
\begin{equation}
	j_\mu = \bar \psi_1 \gamma_\mu \psi_1 .
\label{3d-10}
\end{equation}
Calculations with the spinor \eqref{1-60} give us
\begin{eqnarray}
	j_\mu = \left\lbrace
		\Theta_1^2 + \Theta_2^2 +
		\Theta_3^2 + \Theta_4^2 ,
		r \left( \Theta_1 \Theta_2 - \Theta_3 \Theta_4 \right),
		0,
		\frac{r}{2} \left[
			\sin \theta \left(
				\Theta_1^2 - \Theta_2^2 -
				\Theta_3^2 + \Theta_4^2
			\right) - 2 \cos \theta \left(
				\Theta_1 \Theta_2 - \Theta_3 \Theta_4
			\right)
		\right]
	\right\rbrace .
\label{3d-20}
\end{eqnarray}
Similar result  can be obtained for the spinor $\psi_2$ by changing $\Theta \rightarrow \Sigma$. For the solution \eqref{3c-70}, we have
\begin{equation}
	j_\mu = 2 \tilde \Theta^2 \left\lbrace
		1, \frac{r}{2}, 0, - \frac{r}{2} \cos \theta
	\right\rbrace .
\label{3d-30}
\end{equation}
The Hopf bundle with the metric \eqref{1-15} has the Hopf invariant defined as
\begin{equation}
	H = \frac{1}{V} \int \Upsilon \wedge d \Upsilon~,
\label{3d-40}
\end{equation}
where $\Upsilon$ is some 1-form which can be related to the spatial part of the covariant current \eqref{3d-30}, and
$V = \int \sin \theta d \chi d \theta d \varphi$ is the volume of a $S^3$ sphere of unit radius. We construct the 1-form $\Upsilon$ from the spatial part of the current \eqref{3d-30}:
\begin{equation}
	\Upsilon = \frac{1}{r \tilde \Theta^2} j_i d x^i =
	d \chi - \cos \theta d \varphi,
\label{3d-50}
\end{equation}
where $i = \chi, \theta, \varphi$. Substitution of $\Upsilon$ into \eqref{3d-40} gives us
\begin{equation}
	H = 1 .
\label{3d-60}
\end{equation}
We see that the topological nontriviality of the solution \eqref{3c-70} is connected with the Hopf invariant \eqref{3d-60}.

\subsection{Comparison with the Taub-NUT and Friedman solutions}

Let us compare the solution \eqref{3c-60}-\eqref{3c-80} with the Taub-NUT solution
\begin{equation}
	ds^2 = \frac{dt^2}{U(t)} - 4 l^2 U(t) \left(
		d \chi + \cos \theta d \phi^2
	\right) - \left(
		t^2 + l^2
	\right) \left(
		d \theta^2 + \sin^2 \theta d \phi^2
	\right),
\label{3f-10}
\end{equation}
where $m, l$ are constants and
\begin{equation}
	U(t) = \frac{- t^2 + 2mt + l^2}{t^2 + l^2}.
\end{equation}
The Taub-NUT metric describes an empty spacetime with a nontrivial topology: the spatial section is a 3D sphere $S^3$ that is the total space of the Hopf bundle. 2D space with the metric
$d \theta^2 + \sin^2 \theta d \phi^2$ is the metric on the base of the bundle and the coordinate $\chi$ is spanned on the fibre $S^1$.

We can interpret the solution \eqref{3c-20}, \eqref{3c-30},  \eqref{3c-60}-\eqref{3c-80} as the Taub-NUT spacetime filled with the spinor fields $\psi_{1,2}$ satisfying the Dirac equations \eqref{0-20}.
Physically, this means that the empty nontrivial Taub-NUT spacetime with the Hopf bundle as the spatial section has a nontrivial evolution in time.
But if we fill the spacetime with two spinors $\psi_{1,2}$ and cosmological constant then the spacetime becomes static.

The FRW metric is
\begin{equation}
	ds^2 = dt^2 - a^2(t) dS^2_3,
\label{3f-20}
\end{equation}
where $dS^2_3$ is the metric on a 3D sphere $S_3$. $d S^2_3$ can be written in the standard way
$d S^2_3 = d \psi^2 + \sin^2 \psi \left( d \theta^2 + \sin^2 \theta d \varphi^2 \right) $ or as the metric \eqref{1-15} on the Hopf bundle.

The metric \eqref{3f-20} describes a Friedman Universe filled with  matter. The Universe is not static and evolves from an initial singularity
to the maximum size and then to the final singularity. As we see from the solution \eqref{3c-20}, \eqref{3c-30},  \eqref{3c-60}-\eqref{3c-80},
if the Friedman Universe will be filled with two spinor fields $\psi_{1,2}$ instead the matter plus the cosmological constant then the Universe becomes static without any evolution in time.

\section{Discussion and conclusions}

We have considered Einstein-Dirac gravity with cosmological constant and obtained topologically nontrivial solution for two gravitating Dirac spinors.
We have used two spinors and some special choice of quantum numbers $m,n$  in order to obtain a diagonal energy-momentum tensor.
The diagonal energy-momentum tensor allows us to derive the required solution, which:
\begin{itemize}
	\item is not asymptotically flat;
	\item has no any event horizon;
	\item is topologically nontrivial since it is defined on the Hopf bundle and the Hopf invariant is related to the current of the spinor field;
	\item cannot describe any quantum particle because of nonlinearity of the Einstein-Dirac equations: the spinor cannot be normalized on unity;
	\item can be regarded as the Taub-NUT spacetime filled with the spinor field + $\Lambda$;
	\item can be regarded as a Friedman Universe filled with two spinor fields (instead of matter) without time evolution.
\end{itemize}

\section*{Acknowledgements}

This work was supported by Grant $\Phi.0755$  in fundamental research in natural sciences by the MES of RK. I am very grateful to V. Folomeev for fruitful discussions and comments.

\appendix

\section{Solutions of the nongravitating Dirac equation}
\label{app1}

\subsection{$\Theta_4 = \Theta_3 = 0$, $\mu = 0$}

In this case we have the Dirac equations \eqref{1-80} and \eqref{1-90} in the form
\begin{eqnarray}
	\Theta_1^\prime + \Theta_1 \left(
		\frac{\cot \theta}{2} + n
	\right) +
	\Theta_2  \left(
		\frac{1}{4} - \frac{\Omega}{2} - n \cot \theta - \frac{m}{\sin \theta}
	\right) &=& 0 ,
\label{a1-10}\\
	\Theta_2^\prime + \Theta_2 \left(
		\frac{\cot \theta}{2} - n
	\right) +
	\Theta_1  \left(
		- \frac{1}{4} + \frac{\Omega}{2} - n \cot \theta - \frac{m}{\sin \theta}
	\right) &=& 0 .
\label{a1-20}
\end{eqnarray}
The solution is sought in the form
\begin{equation}
	\Theta_2 = \pm \Theta_1 = C \sin^\alpha \left(
		\frac{\theta}{2}
	\right) \cos^\beta \left(
	\frac{\theta}{2}
	\right) ,
\label{2-10}
\end{equation}
where $C$ is an arbitrary constant. Substituting this into equations \eqref{a1-10}  and \eqref{a1-20}, we have the following solutions:
\begin{eqnarray}
	\frac{\Omega}{2} - \frac{1}{4} &=& \pm n ,
\label{2-30}\\
	\alpha &=& \pm \left( n + m \right) - \frac{1}{2} ,
\label{2-40}\\
	\beta &=& \pm \left( n - m \right) - \frac{1}{2} .
\label{2-50}
\end{eqnarray}
Regular solutions for $\Theta_{1,2}$ and for $\theta \in (0, \pi)$ do exist for
$\alpha \geq 0, \beta \geq 0$. This means that we have the following limitations for the quantum numbers $m,n$:
\begin{eqnarray}
	\pm \left( n + m \right) &\geq& \frac{1}{2} ,
\label{2-110}\\
	\pm \left( n - m \right) &\geq& \frac{1}{2} .
\label{2-120}
\end{eqnarray}

\subsection{$\Theta_1 = \Theta_2 = 0$, $\mu = 0$}

In this case we have the Dirac equations \eqref{1-100} and \eqref{1-110} in the form
\begin{eqnarray}
	\Theta_3^\prime + \Theta_3 \left(
		\frac{\cot \theta}{2} + n
	\right) + \Theta_4
	\left(
		\frac{1}{4} + \frac{\Omega}{2} - n \cot \theta - \frac{m}{\sin \theta}
	\right) &=& 0 ,
\label{a2-10}\\
	\Theta_4^\prime + \Theta_4 \left(
		\frac{\cot \theta}{2} - n
	\right) + \Theta_3
	\left(
		- \frac{1}{4} - \frac{\Omega}{2} - n \cot \theta - \frac{m}{\sin \theta}
	\right) &=& 0 .
\label{a2-20}
\end{eqnarray}
The solution is sought in the form
\begin{equation}
	\Theta_4 = \pm \Theta_3 = C \sin^\alpha \left(
		\frac{\theta}{2}
		\right) \cos^\beta \left(
		\frac{\theta}{2}
	\right) ,
\label{a2-30}
\end{equation}
where $C$ is an arbitrary constant. Substituting this into equations \eqref{1-80}-\eqref{1-110}, we have the following solutions:
\begin{eqnarray}
	\frac{\Omega}{2} + \frac{1}{4} &=& \mp n ,
\label{a2-40}\\
	\alpha &=& \pm \left( n + m \right) - \frac{1}{2} ,
\label{a2-50}\\
	\beta &=& \pm \left( n - m \right) - \frac{1}{2} .
\label{a2-60}
\end{eqnarray}
Once again, for the regularity of the functions $\Theta_{3,4}$ it is necessary to have the following limitations on $m,n$:
\begin{eqnarray}
	\pm \left( n + m \right) &\geq& \frac{1}{2} ,
\label{a2-70}\\
	\pm \left( n - m \right) &\geq& \frac{1}{2} .
\label{a2-80}
\end{eqnarray}

\section{Solutions of the nongravitating Dirac equation with $\mu \neq 0$}
\label{app2}

Substituting $\Theta_{3, 4}$ from \eqref{1-80}  and \eqref{1-90} into \eqref{1-100}-\eqref{1-110}, we have the following set of equations:
\begin{eqnarray}
	\Theta_1''+ \cot \theta \Theta_1' + \frac{\Theta_2^\prime}{2} &+&
	\Theta_1 \left[
		\frac{\cot^2 \theta}{4} - \frac{1}{2 \sin^2 \theta} -
		\left(
			n \cot \theta + \frac{m}{\sin \theta}
		\right)^2 - \left(
			\frac{n}{2} \cot \theta + \frac{1}{2} \frac{m}{\sin \theta}
		\right)
	\right] 
\nonumber \\
	&&
	+\Theta_2 \left(
		- \frac{n}{2} + m \frac{\cot \theta}{\sin \theta} +
		\frac{n}{\sin^2 \theta} + \frac{\cot \theta}{4}
	\right) = \alpha \Theta_1 ,
\label{b-10}\\
	\Theta_2'' + \cot \theta \Theta_2' - \frac{\Theta_1^\prime}{2} &+&
	\Theta_2 \left[
		\frac{\cot^2 \theta}{4} - \frac{1}{2 \sin^2 \theta} -
		\left(
			n \cot \theta + \frac{m}{\sin \theta}
			\right)^2 + \left(
			\frac{n}{2} \cot \theta + \frac{1}{2} \frac{m}{\sin \theta}
		\right)
	\right] 
\nonumber \\
	&&
	+\Theta_1 \left(
		- \frac{n}{2} + m \frac{\cot \theta}{\sin \theta} +
		\frac{n}{\sin^2 \theta} - \frac{\cot \theta}{4}
	\right) = \alpha \Theta_2 ,
\label{b-20}\\
	\alpha &=& \frac{1}{16} + \mu ^2 -
	\frac{\Omega^2}{4} + n^2 .
\label{b-30}
\end{eqnarray}
If we consider equation \eqref{b-10} when $\theta < 0$ then we see that solutions with $\Theta_1(-\theta) = \Theta_2(\theta)$,
$\Theta_2^\prime(-\theta) = - \Theta_1^\prime(\theta)$, and
$\Theta_2^{\prime \prime}(-\theta) = \Theta_1^{\prime \prime}(\theta)$ may exist. Then we can write one nonlocal equation
\begin{equation}
\begin{split}
	\Theta_1''(\theta) + \cot \theta \Theta_1'(\theta) - \frac{\Theta_1^\prime (-\theta)}{2} +&
	\Theta_1(\theta) \left[
	\frac{\cot^2 \theta}{4} - \frac{1}{2 \sin^2 \theta} -
	\left(
	n \cot \theta + \frac{m}{\sin \theta}
	\right)^2 - \left(
	\frac{n}{2} \cot \theta + \frac{1}{2} \frac{m}{\sin \theta}
	\right)
	\right] 
 \\
	&
	+\Theta_1(-\theta) \left(
	- \frac{n}{2} + m \frac{\cot \theta}{\sin \theta} +
	\frac{n}{\sin^2 \theta} + \frac{\cot \theta}{4}
	\right) = \alpha \Theta_1(\theta),
\label{b-35}
\end{split}
\end{equation}
and the same for $\Theta_2$. Numerical investigations confirm this statement. 

Another way to solve equations \eqref{b-10} and \eqref{b-20} is the decomposition of $\Theta_{1,2}$ on odd and even functions 
\begin{eqnarray}
	Y_1 &=& \Theta_1 + \Theta_2 \text{ is even function} , 
\label{b-40}\\
	Y_2 &=& \Theta_1 - \Theta_2  \text{ is odd function} .  
\label{b-50}
\end{eqnarray}

We will solve equations \eqref{b-10} and \eqref{b-20} as an eigenvalue problem with the eigenvalue $\alpha$ and eigenfunctions
$\Theta_{1,2}$.

\subsection{Numerical solution for \eqref{b-10} and \eqref{b-20}}

\begin{figure}[h!]
	\fbox{
		\includegraphics[width=.4\linewidth]{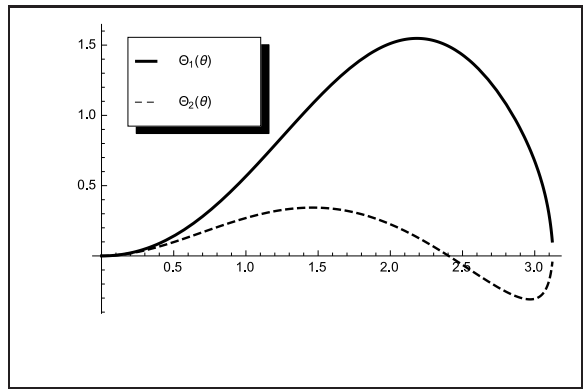}
	}
	\caption{The profiles of $\Theta_{1,2}(\theta)$ for equations \eqref{b-10} and \eqref{b-20}.
	}
\label{fig1}
\end{figure}

In order to avoid a singularity in equations \eqref{b-10} and \eqref{b-20} for the angle $\theta = 0$, we have to use the following form of the functions $\Theta_{1,2}$:
\begin{eqnarray}
	\Theta_1 &=& \frac{\alpha_2}{2} \theta^2 + \frac{\alpha_3}{6} \theta^3 + \ldots ,
\label{b1-10}\\
	\Theta_2 &=& \frac{\beta_2}{2} \theta^2 + \frac{\beta_3}{6} \theta^3 + \ldots 
\label{b1-20}
\end{eqnarray}
Substituting \eqref{b1-10} and \eqref{b1-20} into \eqref{b-10} and \eqref{b-20}, we have
\begin{eqnarray}
	\beta_2 &=& \pm \alpha_2 ,
\label{b1-30}\\
	\beta_3 &=& \pm \alpha_3 ,
\label{b1-40}\\
	m +n &=& \pm \frac{3}{2}, \pm \frac{5}{2} .
\label{b1-50}
\end{eqnarray}
Numerical solutions are presented in Fig.~\ref{fig1} for the following values of the parameters:
\begin{eqnarray}
	m + n &=& \frac{5}{2} , n = 1,
\label{b1-60}\\
	\beta_2 &=& \alpha_2 = 1.0 ,
\label{b1-70}\\
	\beta_3 &=& - \alpha_2 = - 1.0 .
\label{b1-80}
\end{eqnarray}
The numerical calculations give us
\begin{equation}
	\alpha = \frac{1}{16} + \mu ^2 -
	\frac{\Omega^2}{4} + n^2 \approx -4.13208 .
\label{b1-90}
\end{equation}

\end{document}